\documentclass[letterpaper, 10 pt, conference]{ieeeconf}  % Comment this line out if you need a4paper
\usepackage{amsmath,graphicx}
\usepackage[dvipsnames]{xcolor}
\usepackage[hidelinks]{hyperref}
\usepackage{titlesec}
 \usepackage{multirow}
\usepackage{subfigure}
\IEEEoverridecommandlockouts                              % This command is only needed if 
\title{\LARGE \bf A Hybrid Gaze–Motor Imagery BCI Framework for Effective Decision Communication
}

\author{Gowtham Reddy N$^{1}$, KongFatt Wong-Lin$^{2}$ and Yogesh Kumar Meena$^{1}$% <-this % stops a space
\thanks{$^{1}$Gowtham and Yogesh Kumar Meena are with Human-AI Interaction
(HAIx) Lab, IIT Gandhinagar, India.
        {\tt\small yk.meena@iitgn.ac.in}}%
\thanks{$^{2}$KongFatt Wong-Lin with Intelligent Systems Research Centre (ISRC), Ulster University, UK.
        }%
}

\begin{document}

\maketitle
\thispagestyle{empty}
\pagestyle{empty}

%%%%%%%%%%%%%%%%%%%%%%%%%%%%%%%%%%%%%%%%%%%%%%%%%%%%%%%%%%%%%%%%%%%%%%%%%%%%%%%%
\begin{abstract}

Non-invasive brain-computer interfaces (BCIs) and eye-tracking technologies offer promising communication pathways; however, motor imagery (MI)-based BCIs often suffer from low discriminability and high inter-subject variability. To mitigate these issues, this study investigates the impact of visual fixation on neural response stability in both standalone MI and hybrid MI–eye tracking systems. We then propose a novel asynchronous hybrid paradigm that streamlines user intent by utilising eye-tracking for direct selection, followed by MI-based confirmation, significantly reducing the operational steps required by conventional systems. The paradigm was evaluated with 15 healthy participants using a 16-channel EEG system.
Results show that MI-related information is predominantly localised within motor cortex regions, with limited-channel configurations (SVM: 0.58) achieving performance comparable to full-montage setups (SVM: 0.54). 
The hybrid MI paradigm further outperforms conventional MI, achieving up to 100\% accuracy with greater robustness across all channel configurations.
Our findings indicate that visual fixation enhances neural response stability, while integrating eye-tracking with MI enables the development of reliable, scalable multi-command BCI systems suitable for real-world applications.

% Overall, fixation stabilises neural responses, and integrating eye-tracking with MI enables more reliable and scalable multi-command BCI systems for real-world applications.

\end{abstract}

% \begin{IEEEkeywords}
% EEG, Motor Imagery, Hybrid BCI .
% \end{IEEEkeywords}
%%%%%%%%%%%%%%%%%%%%%%%%%%%%%%%%%%%%%%%%%%%%%%%%%%%%%%%%%%%%%%%%%%%%%%%%%%%%%%%%
\section{INTRODUCTION}
Brain–computer interfaces (BCIs) enhance effective decision communication by decoding neural signals, such as electrophysiological activity and slow cortical potentials, into actionable commands or messages, thereby bypassing traditional motor pathways \cite{mcfarland2011brain}.
Recent advancements in \textcolor{black}{BCI} systems have enabled efficient interaction between humans and machines through neural signals, with motor imagery (MI) based paradigms emerging as one of the most widely explored approaches for decoding user intent from electroencephalogram (EEG) signals. 
Building on this foundation, several studies have proposed various MI-based paradigms and signal-processing approaches to improve classification performance and system reliability \cite{Pfurtscheller, tangermann2012review}.

Despite these advancements, MI-based BCI systems continue to face several challenges, including low signal-to-noise ratio (SNR) and high inter-subject variability. To address these limitations, hybrid BCI approaches have emerged as a promising direction by integrating neurological signals (e.g., MI, SSVEP, P300) with other physiological signals such as eye-tracking \cite{10.3389/fnpro.2010.00003, meena2017hybrid, tan2022autonomous}. By leveraging these additional modalities, hybrid systems provide supplementary and often more stable information, thereby reducing reliance on noisy EEG signals. Among these, eye tracking provides a natural and intuitive means of interaction, offering fast and reliable target selection that can complement the relatively slower and variable MI responses. 
O’Doherty et al. \cite{o2014exploring} explored the design of a gaze–MI hybrid BCI, demonstrating the feasibility of integrating eye gaze with MI to improve interaction efficiency. Building on this, Meena et al. \cite{meena2015towards} proposed a hybrid BCI interface that combines gaze and MI to increase the number of commands, thereby improving interaction efficiency.

Further extending this concept, Meena et al. \cite{meena2017hybrid} introduced a hybrid BCI interface combining eye-tracking and MI to enable faster sequential decision-making using a standard interface layout across multiple input modes. The system demonstrated improved efficiency, including an approximate 45\% reduction in MI effort, highlighting its potential for clinical and psychological applications. Zhang et al. \cite{zhang2024high} proposed a hybrid MI and eye-tracking-based BCI system for virtual cursor control, achieving high accuracy and stable interaction for performing common computer tasks.

Despite these promising developments, most existing hybrid BCI systems are predominantly based on \textcolor{black}{cue-paced (synchronous)} paradigms, where user interactions are constrained by predefined cues and fixed time windows for decoding left- and right-hand motor imagery. In contrast, relatively few studies have explored \textcolor{black}{self-paced (asynchronous)} hybrid BCI frameworks \cite{lotte2008self, li2013hybrid}, particularly those employing one-versus-rest motor imagery strategies that function as a brain switch for continuous and self-paced control. In such paradigms, instead of requiring users to continuously switch between multiple motor imagery tasks (e.g., left vs. right), a single MI class (e.g., right-hand imagery) can act as a control or “selection” command, while the rest/idle state serves as a non-control or “waiting” state, thereby reducing cognitive load. 
% This limitation restricts the practical usability and flexibility of hybrid systems in real-world scenarios, where faster interaction and reduced cognitive demand are essential compared to synchronous approaches. 
Existing studies predominantly investigate perceptual decision-making, gaze behaviour, and MI-based BCI paradigms in isolation, with limited efforts toward their unified integration. Although eye-tracking and fixation mechanisms provide valuable cognitive and behavioural insights, their potential to enhance real-time MI-based BCI performance remains underexplored. 
% Consequently, the absence of a cohesive hybrid framework integrating self-paced MI with eye-tracking constrains system robustness, adaptability, and ecological validity.

% In particular, self-phased hybrid BCI frameworks have gained increasing attention, as they enable users to operate the system in a self-paced manner without relying on strict cue-based paradigms, thereby improving flexibility and decision-making efficiency.
Motivated by these considerations, this work investigates integrating self-phased MI with eye-tracking signals to develop a more robust and efficient hybrid BCI paradigm \cite{meena2015simultaneous}.
In designing the proposed paradigm, we draw upon insights from prior neurocognitive and BCI research. Studies in perceptual decision-making \cite{ shooshtari2019confidence, heekeren2008neural} have shown that sensory (e.g. visual motion) stimuli play a critical role in guiding decisions, where the brain accumulates sensory evidence to inform a choice and corresponding action. 
% from dynamic stimuli to support intentional actions.
Building on this principle, Basteris et al. \cite{basteris2021epsychology} \textcolor{black}{employed an online}
coherent-motion dot task to investigate perceptual decision-making, demonstrating that eye-movement patterns can effectively reflect decision accuracy. Furthermore, Krause et al. \cite{krause2023maintaining} showed that maintaining gaze fixation improves the resolution of spatial cognitive conflicts and stabilises cognitive control without degrading task performance. Complementary findings by Bergeron et al. \cite{bergeron2000fixation} and Munoz et al. \cite{munoz2002vying} highlight the functional role of fixation-related neural mechanisms, indicating that fixation and saccade neurons dynamically regulate goal-directed visual behaviour. In the context of MI-based BCI, Meng et al. \cite{meng2022effects} demonstrated that system performance remains robust across different gaze-fixation conditions, suggesting that users can operate MI-BCI systems without strict fixation constraints. 

Building on these findings, we design a novel perceptual decision-making task paradigm that integrates MI with visual attention mechanisms, incorporating both fixation and non-fixation conditions. The main contributions of this work include: (1) designing an attention-aware MI and hybrid BCI paradigm incorporating fixation, peripheral targets, colour cues, and dot-motion feedback to modulate attention; (2) \textcolor{black}{ developing an asynchronous hybrid MI–eye-tracking framework enabling self-paced (one-vs-rest, brain-switch-style architecture), low cognitive-load decision-making;} and (3) conducting a comprehensive comparison with conventional MI, and hybrid MI demonstrating improved performance and efficiency.

% \item A novel attention-aware spatial MI and hybrid MI-based BCI paradigm is introduced, integrating central fixation, peripheral target arrangement, color-coded cueing, and \textcolor{black}{dot-motion–based visual feedback to effectively modulate visual attention}; additionally, a comprehensive evaluation is conducted under fixation-constrained (covert attention) and free-gaze (overt attention) conditions to analyze their impact on BCI performance.

%     \item An asynchronous hybrid BCI framework integrating MI with eye tracking is developed, enabling self-paced control through a one-versus-rest strategy, thereby reducing cognitive load and improving decision-making.
    
%     \item A detailed comparative analysis between conventional MI-based BCI and the proposed hybrid MI–eye-tracking BCI is performed to evaluate improvements in decision-making performance and system efficiency.
    
%     % \item The proposed hybrid paradigm significantly improves interaction efficiency by enabling direct single-step selection through eye tracking with asynchronous MI-based confirmation, in contrast to conventional MI systems that require multiple sequential executions for a single command.
% \end{itemize}

 \section{Materials and Methods}
 \subsection{Participant Demographics}
 15 healthy participants (20-26 years, 14 male and 1 female) were enrolled in this study. Participants were recruited voluntarily, and all experiments were conducted under controlled laboratory conditions, following a standardised data acquisition protocol to ensure consistency across trials. \textcolor{black}{The study was reviewed and approved by the Institutional Ethics Committee of the Indian Institute of Technology Gandhinagar, India, and adhered to standards outlined in the Declaration of Helsinki.}

\subsection{Experimental Protocol}
Prior to the start of each experimental session, the configuration parameters are defined, including the input mode, fixation condition (on or off), operational phase (calibration or testing), and the number of cycles to be executed. One cycle includes 24 trials in the MI task and 8 trials in the hybrid eye-tracking and MI paradigm. The trials are then conducted under two fixation conditions with a fixation cross (+) and without a fixation cross. The fixation-cross condition supports attentional alignment and controlled readiness, whereas the no-fixation condition is used to evaluate how participants self-regulate attention when no external stabilisation reference is provided. In both conditions, the dot-based visualisation provides progressive feedback during the decision process, allowing participants to monitor the stability of their responses and adjust their actions before final confirmation, thereby supporting accurate and intentional target selection. Building upon this interaction framework, effective decision communication is examined through the design and implementation of MI and hybrid (eye-tracking and MI) BCI paradigms, as described below.

\subsubsection{MI Paradigm Design}
\begin{figure}
\centering
\includegraphics[width=0.71\linewidth]{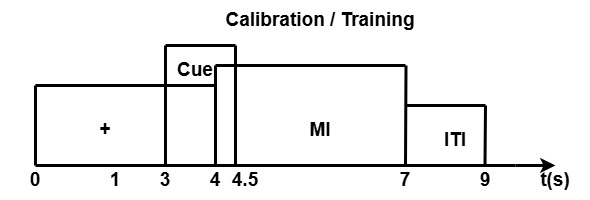}
\caption{Timing diagram of each trial of the MI / Hybrid signals task.}
\label{timing_mi}
\end{figure}

\begin{figure}
  \centering
  \subfigure[Fixation]{\includegraphics[width=0.2\textwidth]{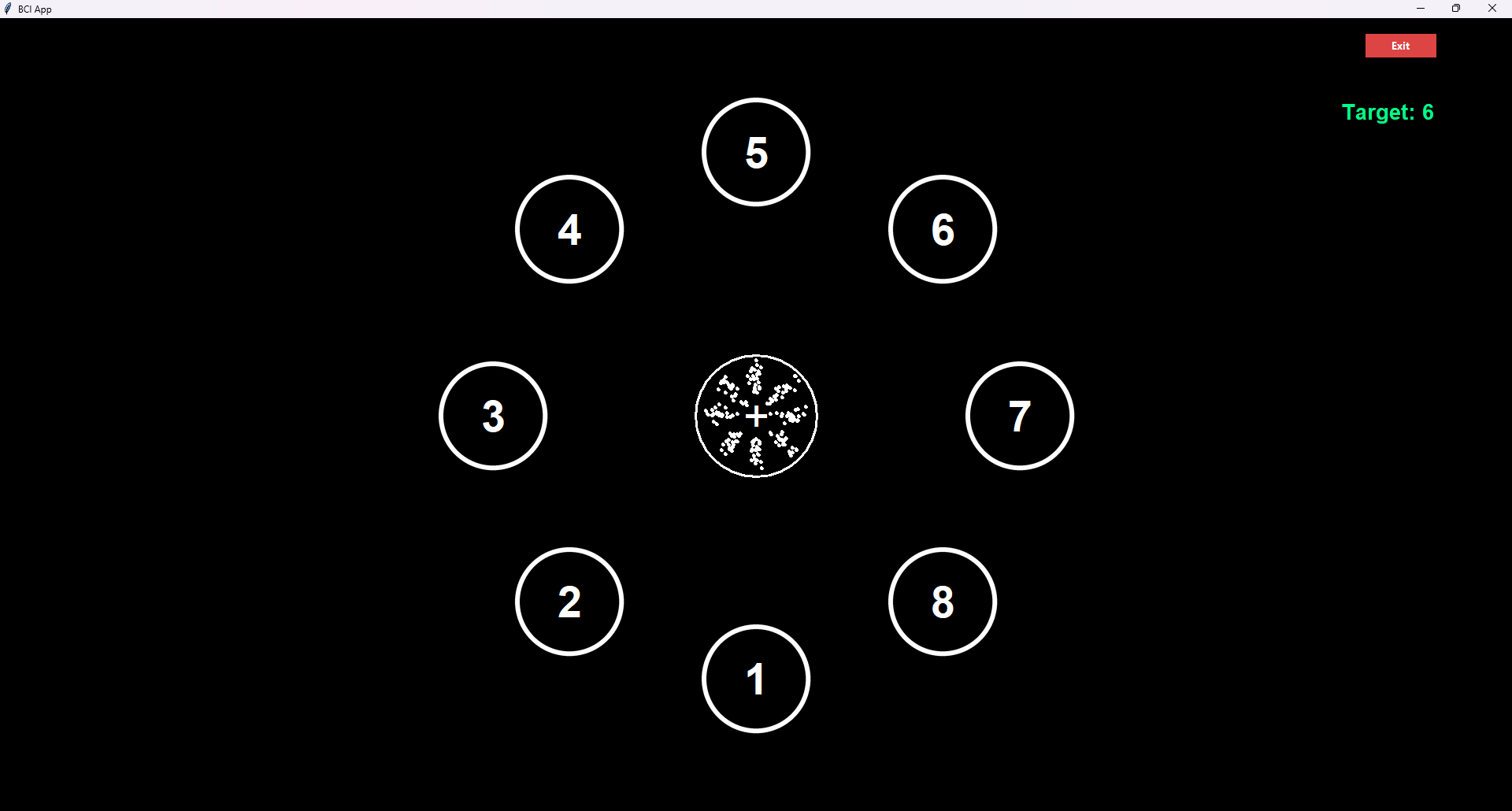}}
  % \qquad
  \subfigure[Cue]{\includegraphics[width=0.2\textwidth]{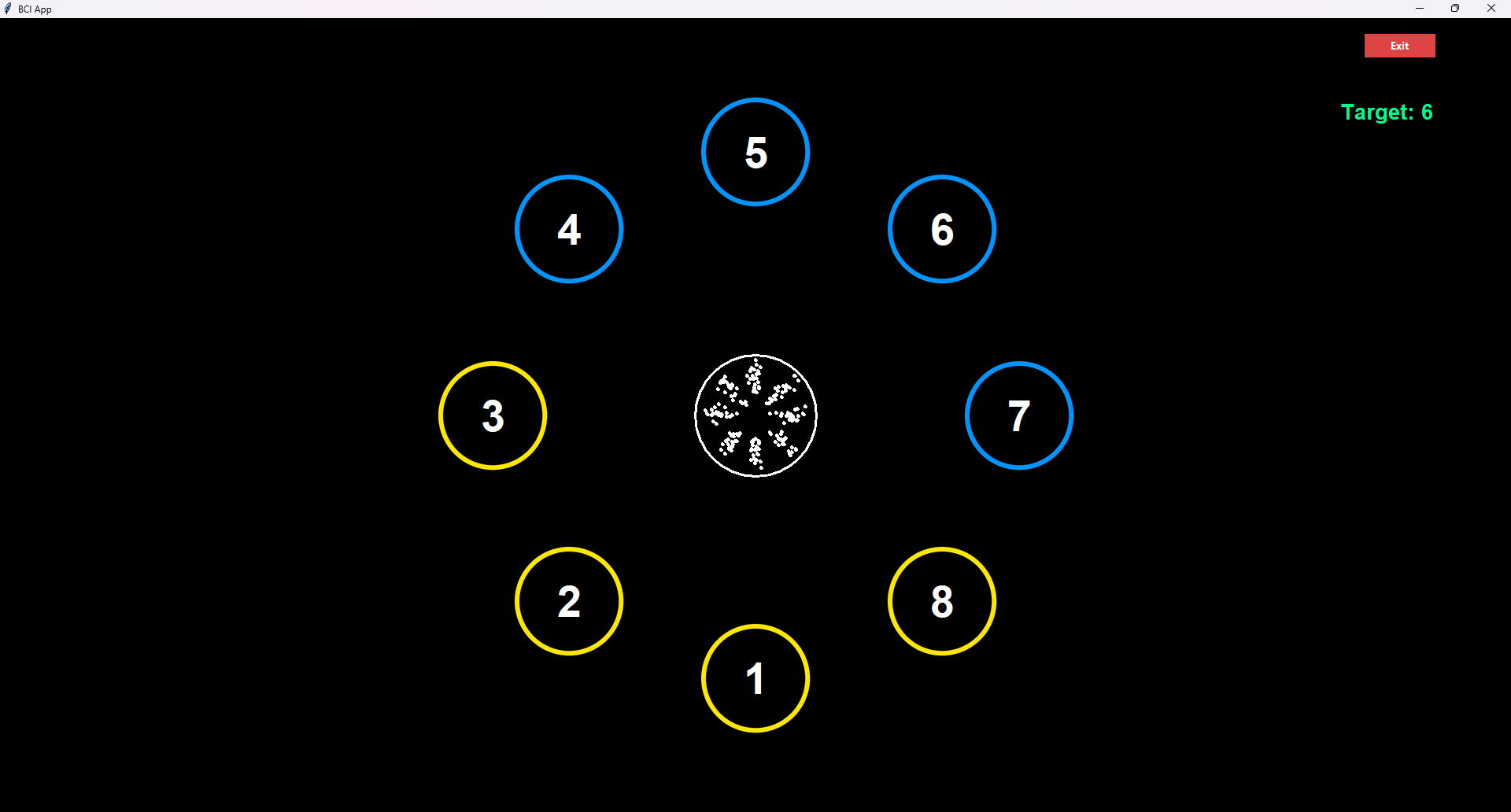}}
  \qquad
  \subfigure[First Decision]{\includegraphics[width=0.2\textwidth]{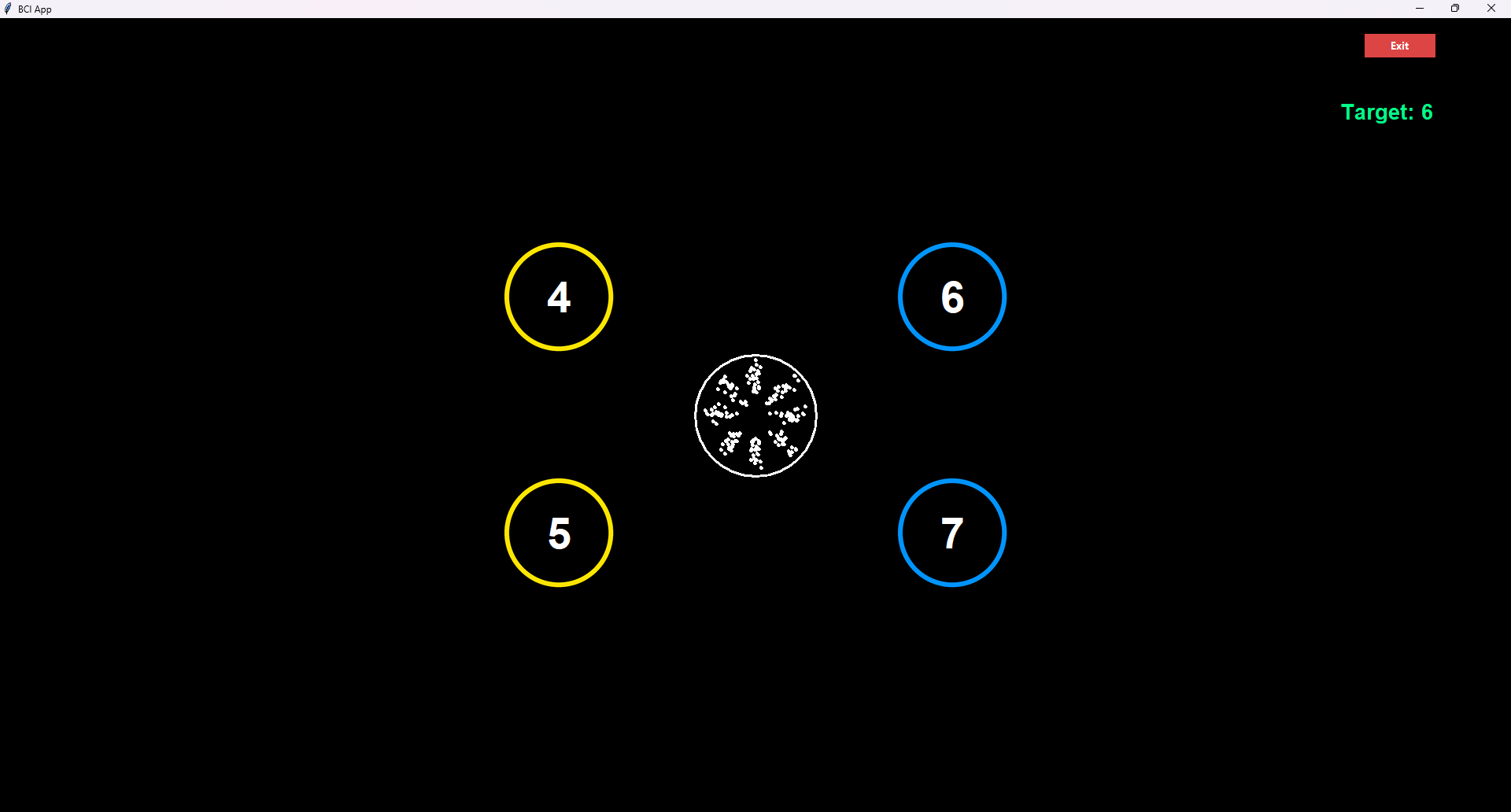}}
  % \qquad
  \subfigure[Second Decision]{\includegraphics[width=0.2\textwidth]{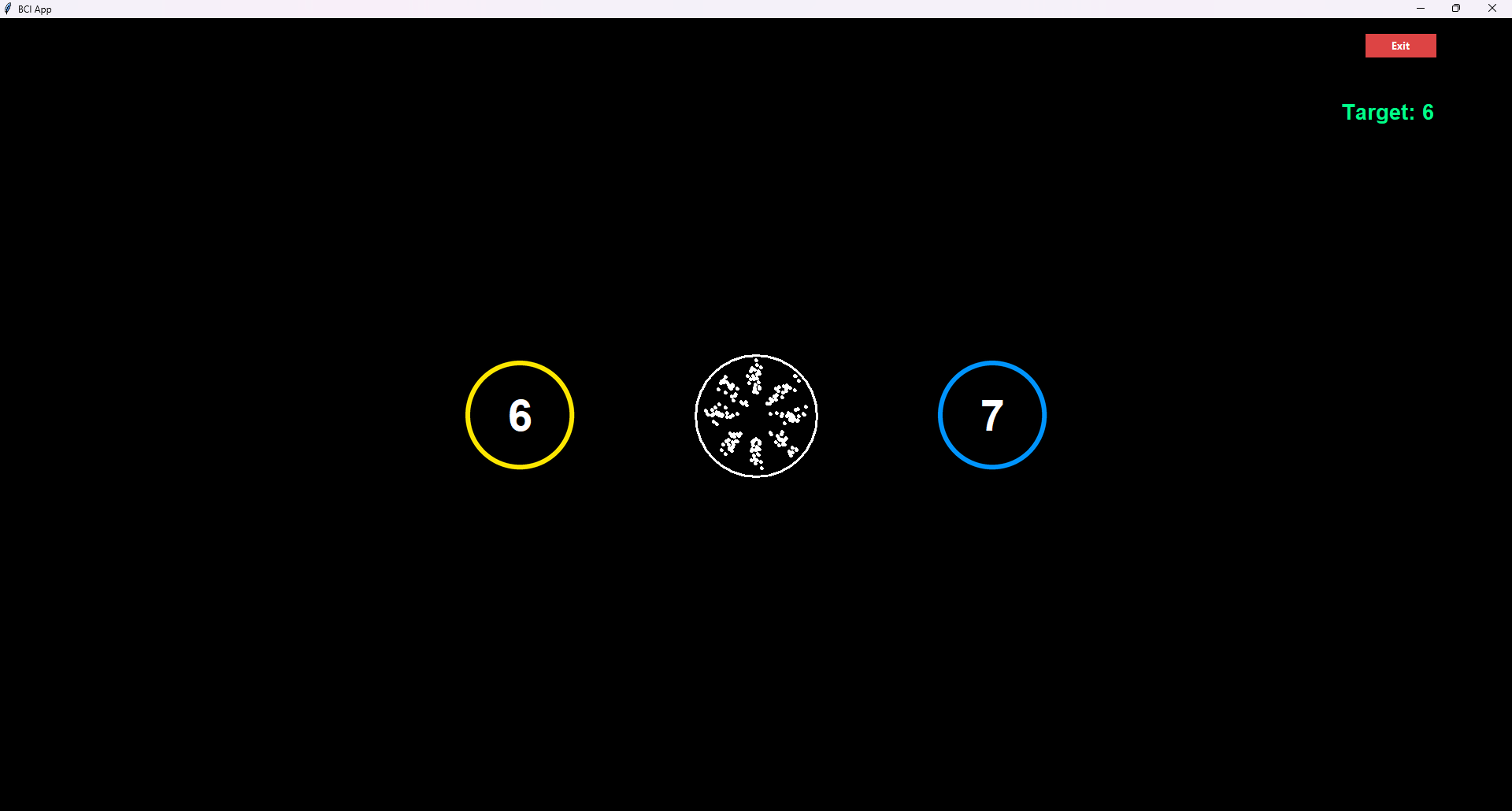}}
  \caption{{Illustration of the proposed MI-based target selection paradigm showing (a) fixation, (b) colour-coded cue presentation, and (c–d) target-driven hierarchical decision stages.}}
  \label{fig:single MI Cycle}
\end{figure}

The MI paradigm is designed to capture the neural representation of intended movements by decoding motor imagery-related brain activity, thereby enabling reliable identification of participants’ action intentions (Fig. ~\ref{timing_mi}). To ensure a direct correspondence between neural activity and the visual interface, the MI task is mapped onto a circular layout consisting of eight uniformly distributed targets, associated with a distinct spatial cue (Fig. \ref{fig:single MI Cycle}(a)).
To facilitate efficient target selection, the framework uses a \textcolor{black}{hierarchical decision strategy in which the eight targets are progressively narrowed down through binary splits (8 → 4 → 2 → 1)}. Initially, the eight targets are divided into two groups of four, each represented by a different colour. The participant first selects one group (e.g., left or right) \textcolor{black}{in the direction of the perceived motion coherent direction (Fig. \ref{fig:single MI Cycle}(b))}. In the next stage, the selected group of four is further divided into two smaller groups of two, and the participant \textcolor{black}{makes another selection, based on the perceived motion coherence direction (Fig. \ref{fig:single MI Cycle}(c))} again makes a selection. This \textcolor{black}{decision tree} process continues, splitting the options at each step, until a single target is finally selected (Fig \ref{fig:single MI Cycle}(d)). This step-by-step refinement enables accurate and structured decision-making.

\textcolor{black}{Each calibration trial begins with a 3-second fixation cross (+) presented at the centre of the screen, stabilising the participant and minimising motion artefacts. Following this fixation period, a visual cue appears at a random onset within the 3–3.5 second window and persists for 1 second, triggering the participant to initiate the instructed MI task. The participant then executes the corresponding MI task over a 2.5-second response window. Finally, a 2-second inter-trial interval (ITI) allows the participant to rest before the next trial commences.}
% Each calibration trial begins with a 3-second central fixation cross to stabilize the participant and reduce motion artifacts. A motor imagery (MI) cue is then presented for 1 second, after which the participant performs the corresponding MI task during a 3-second response period. Finally, a 2-second inter-trial interval allows the participant to relax before the next trial.
The overall timing structure is shown in Fig. \ref{timing_mi}, and a complete MI cycle is illustrated in Fig. \ref{fig:single MI Cycle}.

\subsubsection{Hybrid Eye-tracking and MI Paradigm}
\begin{figure}
  \centering
  \subfigure[Fixation]{\includegraphics[width=0.425\linewidth]{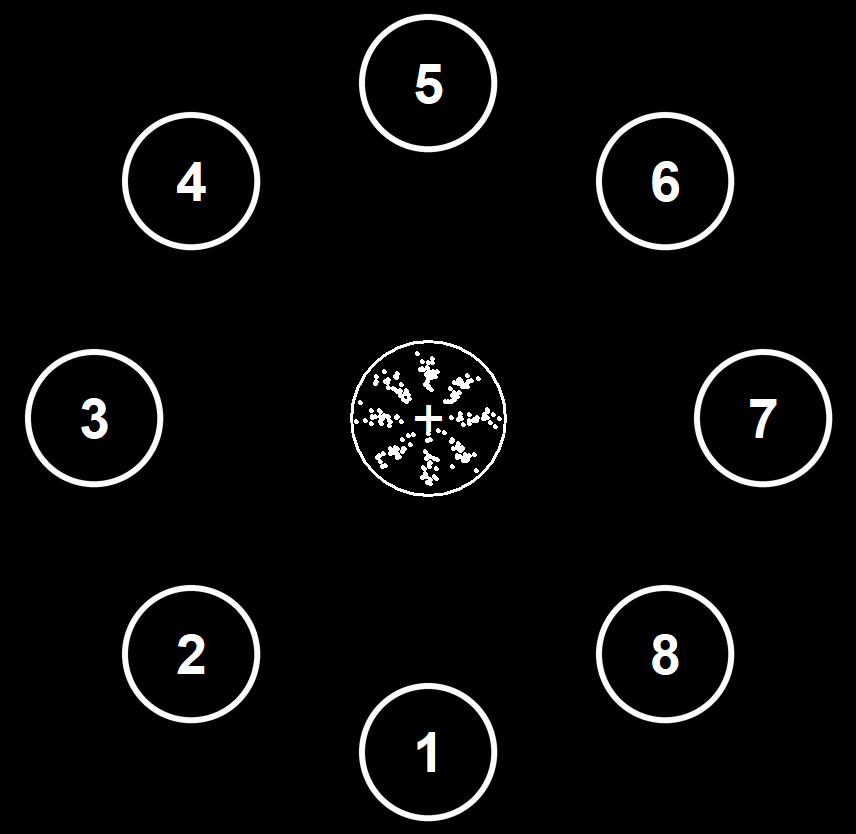}}
  \qquad
  \subfigure[Cue]{\includegraphics[width=0.44\linewidth]{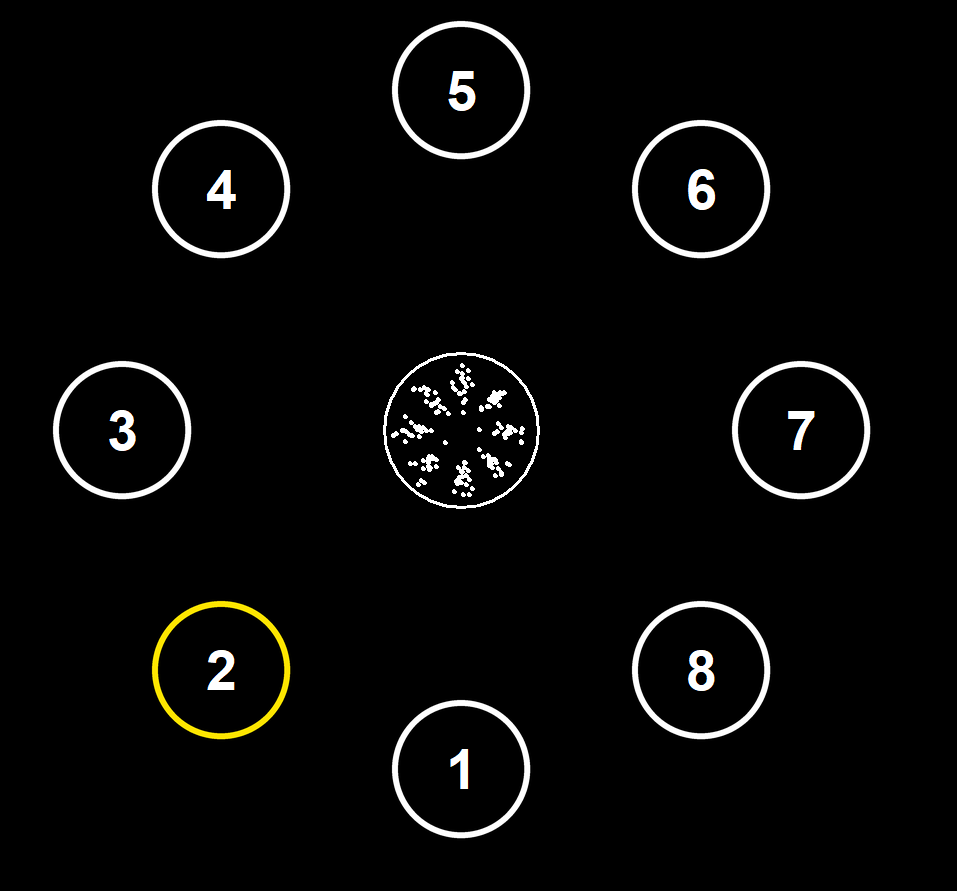}}
  % \qquad
  % \subfigure[Decision]{\includegraphics[width=0.412\linewidth]{Decision.png}}
  \caption{{Illustration of a single hybrid trial showing fixation (left), colored cue (right).}}
  \label{fig:hybrid MI Cycle}
\end{figure}

The hybrid BCI paradigm follows the same trial timing structure as described in the MI paradigm (fixation cross, colored cue, decision phase, and ITI), with the addition of synchronised eye-tracking. During the decision phase, the participant fixates on the cued target while simultaneously performing the corresponding motor imagery task.
% Figure \ref{hybridtiming} presents the complete timing diagram of a single trial within the hybrid BCI paradigm.
% The hybrid BCI paradigm integrates eye-tracking-based target selection with MI to enable a unified and decision-making process. Each hybrid trial follows a structured timing sequence that combines both modalities within a single trial flow. The calibration trial begins with a 3-second start screen, followed by a 1-second central fixation cross to stabilise gaze and prepare the participant for the upcoming tasks. A 3-second decision phase is then presented, during which the participant must fixate on the highlighted target and simultaneously performing the cued MI task.
% % Within this interval, a 1.5-second segment of both eye-gaze and EEG activity is recorded for analysis, ensuring stable fixation and consistent motor-task execution. 
% A 2-second inter-trial interval is provided at the end of each trial to allow the participant to relax before the next sequence begins.
% Figure \ref{hybridtiming} illustrates the full timing of the single-trial hybrid paradigm. 
Fig. \ref{fig:hybrid MI Cycle} illustrates the experimental procedure of the hybrid BCI paradigm. 
% Each trial begins with a fixation phase, followed by a color-coded cue indicating the target. During the decision phase, the user looks at the cued target while simultaneously performing the corresponding motor imagery task. 
For example, if the cue indicates target 2, the user directs their gaze to target 2 and performs the associated motor imagery task (e.g., right-hand imagery). The specific target and motor imagery task may vary across trials. This combined action continues until the rest phase.
% Figure \ref{fig:hybrid MI Cycle} illustrates the experimental procedure for the hybrid BCI paradigm. Each trial begins with a fixation phase, followed by the presentation of a color-coded cue. During the decision window, the user is required to shift their gaze toward the cued target (e.g., target 2 as hown in figure) while simultaneously performing the corresponding motor imagery task (e.g., right-hand imagery). This combined eye movement and MI process is sustained until the rest phase, completing a single trial.

\subsection{Experimental Task}
In the MI paradigm, participants were first instructed on how to perform the task. At the start of each trial, a target was presented. In the fixation condition, subjects were required to maintain gaze on a central fixation cross (+), whereas in the non-fixation condition, they were free to look anywhere on the screen. Following this, based on the timing diagram, the paradigm transitioned into cue presentation, where left or right directional cues were displayed. \textcolor{black}{Then the subjects performed the corresponding} MI as per the target location (left or right) during the decision phase until the target selection period was completed.
In the hybrid MI–eye-tracking paradigm, the fixation and cue presentation followed the same protocol as in the MI paradigm. During the activity period, subjects were instructed to simultaneously gaze at the target and perform the corresponding right-hand MI. This integration of eye gaze and MI was designed to enhance user engagement and improve classification performance. A demonstrative video of the complete experimental workflow is available online~\footnote{\url{https://youtu.be/K5K3pO6PDpE}}

% \begin{figure}
% \centering
% \includegraphics[width=0.71\linewidth]{Hybrid MI_new.jpg}
% \caption{ Timing diagram of the hybrid BCI task.}
% \label{hybridtiming}
% \end{figure}

\subsection{Data Acquisition}
The EEG data were acquired \textcolor{black}{for 9 trials per task} using the g.Nautilus research-grade wireless headset equipped with 16 electrodes, with a sampling frequency of 500 Hz. The EEG cap was carefully positioned on each participant according to the standard  10–20 electrode placement protocol, and conductive gel was applied to ensure low electrode-skin impedance and high-quality signal acquisition. Prior to recording, impedance levels were verified and maintained within acceptable limits to minimise noise and artefacts. EEG signals were continuously recorded throughout the entire trial for both the MI and hybrid paradigms, following the sequence and timing defined in the timing diagram. Labels were assigned according to this timing structure, ensuring that each EEG segment corresponded to its respective phase (e.g., fixation, cue, decision, and rest) and task condition.

In MI paradigm, the continuous EEG data were segmented into trials based on changes in the class label (left and right motor imagery). A block-wise segmentation approach was adopted, where consecutive samples with identical labels were grouped into a single trial. Each trial was truncated to 1500 samples to ensure uniform length across all epochs. while in the hybrid MI paradigm, the continuous EEG data were segmented into epochs based on event markers corresponding to `right-hand MI’ and `rest’ phases. Each segment was identified by detecting transitions in the event labels, and fixed-length epochs were extracted accordingly. For the MI (decision) phase, a time window of 500 to 1500 samples was selected to capture task-relevant neural activity, yielding 1000-sample epochs. For the rest phase, the first 1000 samples of each segment were extracted. This segmentation approach ensures consistent temporal alignment across trials while effectively isolating neural patterns associated with motor imagery and idle states.  For both the MI and hybrid MI paradigms, the analysis was conducted across three channel configurations: motor cortex channels (C3, Cz, C4), extended motor cortex (F3, Fz, F4, P3, Pz, P4 and motor cortex channels), and the full 16-channel EEG montage. This setup enables a comprehensive evaluation of how spatial channel selection influences classification performance, while also assessing the contribution of additional neighbouring electrodes around the motor cortex.
% Two channel configurations were considered for analysis:
% Motor cortex channels (C3, Cz, C4) and Full 16-channel EEG montage. 

\subsection{Preprocessing}
 To isolate task-relevant neural oscillations associated with MI, each segmented trial was then decomposed into two physiologically significant frequency bands: the mu band (8–12 Hz) and the beta band (13–30 Hz), which are well known to reflect ERD and ERS in the sensorimotor cortex. A fourth-order Butterworth bandpass filter was employed to extract these frequency components, owing to its maximally flat frequency response in the passband. 
 % Forward–backward filtering was applied to achieve zero-phase distortion, thereby preserving the temporal characteristics of the EEG signals. 
 The filtering process was performed independently for each trial and across different channel configurations.
Subsequently, artefact removal was carried out using Fast Independent Component Analysis (ICA), where components highly correlated with frontal channels (Fp1, Fp2) were identified and removed to suppress ocular artefacts. 

\subsection{Feature Extraction using CSP}

Spatial filtering was performed using the CSP \cite{ramoser2000optimal} algorithm to enhance class separability. CSP computes spatial filters that maximise the variance for one class while minimising it for the other, thereby improving discriminative feature extraction for MI tasks.
For each trial, the normalised spatial covariance matrix was computed. 
% as:
% \begin{equation}
% \mathbf{C} = \frac{\mathbf{X}\mathbf{X}^T}{\mathrm{trace}(\mathbf{X}\mathbf{X}^T)}
% \end{equation}
% where $\mathbf{X} \in \mathbb{R}^{N \times T}$ represents the EEG data matrix with $N$ channels and $T$ time samples.
% The average covariance matrices for the two classes were denoted as $\mathbf{R}_1$ and $\mathbf{R}_2$. A composite covariance matrix was formed and decomposed using a generalized eigenvalue problem:
% \begin{equation}
% \mathbf{R}_1 \mathbf{w} = \lambda (\mathbf{R}_1 + \mathbf{R}_2)\mathbf{w}
% \end{equation}
% where $\lambda$ represents the eigenvalues and $\mathbf{w}$ denotes the spatial filters.
% The eigenvectors corresponding to the largest and smallest eigenvalues were selected to construct the CSP projection matrix, as they capture the most discriminative variance patterns between the two classes.
% Each EEG trial was then projected onto the selected spatial filters, and the log-variance of the filtered signals was computed to form feature vectors:
% \begin{equation}
% f_i = \log \left( \frac{\mathrm{var}(\mathbf{Z}_i)}{\sum_{j} \mathrm{var}(\mathbf{Z}_j)} \right)
% \end{equation}
% where $\mathbf{Z} = \mathbf{W}^T \mathbf{X}$ represents the spatially filtered signal.
For each frequency band, four CSP components were selected. The final feature vector was obtained by concatenating features from both the mu and beta bands, resulting in an dimensional feature representation per trial.

\subsection{Classification}
The extracted features were evaluated using multiple supervised machine learning classifiers, including Support Vector Machine (SVM) with a radial basis function kernel, Random Forest (RF), Decision Tree (DT), K-Nearest Neighbours (KNN), and Gaussian Naïve Bayes (NB). The dataset was evaluated using 5-fold cross-validation, in which the data were partitioned into five subsets, with each subset used once as the test set and the remaining four as the training set. To ensure transparency and reproducibility, source code is available at the Hybrid-BCI-for-Effective-Decision-Communication repository~\footnote{\url{https://github.com/HAIx-Lab/Hybrid-BCI-for-Effective-Decision-Communication}}

% and dataset details are publicly available at: \href{https://github.com/Gowthamgit04/Hybrid-Gaze-MI-BCI.git}{Hybrid-Gaze-MI-BCI}.

% All source code and dataset details are available at the Multimodal-Gaze-Semantic-Framework repository~\footnote{\url{https://github.com/HAIx-Lab/Multimodal-Gaze-Semantic-Framework}} to ensure reproducibility of this work.

\section{Results}

\begin{figure*}[ht]
\centering
\includegraphics[width=1\textwidth]{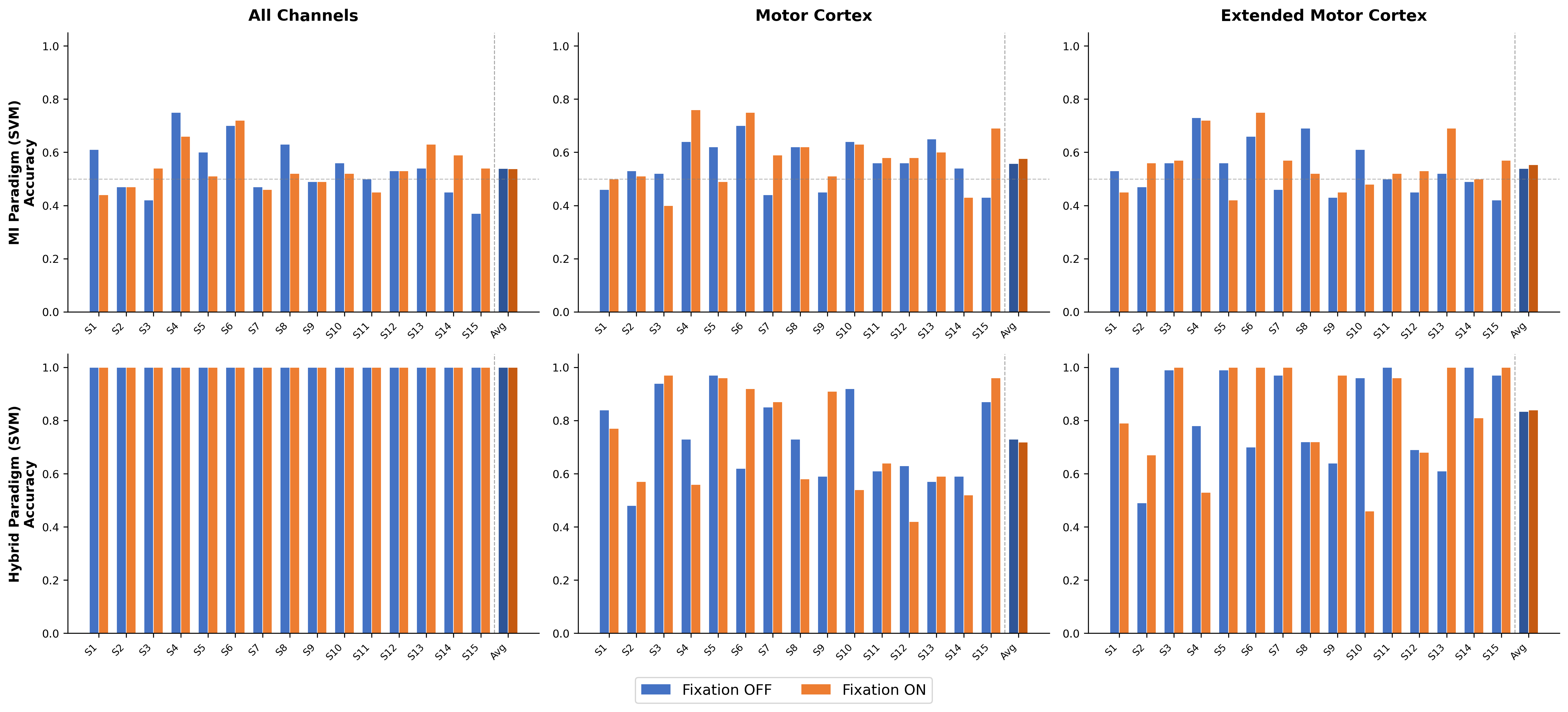}
\caption{Subject-wise classification accuracies (SVM) for the MI paradigm (top row) and the hybrid MI paradigm (bottom row) across three channel configurations: all channels (left), strict motor cortex (centre), and extended motor cortex (right), under fixation-off and fixation-on conditions. }
\label{fig:barplot_mi_hybrid}
\end{figure*}
% The rightmost bar in each subplot represents the mean accuracy across all subjects.

The classification performance of the conventional and hybrid MI paradigms under different EEG channel configurations is illustrated in \textcolor{black}{Fig.~\ref{fig:barplot_mi_hybrid}, which (top) presents classification accuracies for the conventional MI paradigm and the hybrid MI paradigm across three channel configurations (all channels, strict motor cortex, and extended motor cortex) under fixation-off and fixation-on conditions.} During MI (top row), all-channel configuration: Under the fixation-off condition, mean accuracies ranged from 0.52 (NB) to 0.55 (RF, DT), with standard deviations of approximately 0.11 across classifiers. Under fixation-on, mean accuracies remained comparable (0.53–0.55), but inter-subject variability showed a modest reduction (SD: 0.08–0.11), indicating that fixation contributes to more stable neural responses without dramatically changing overall mean accuracy. While the strict motor cortex configuration yielded slightly higher mean accuracy under the fixation-on condition (SVM: 0.58, SD: 0.11) compared to fixation-off (SVM: 0.56, SD: 0.08), suggesting that focused motor-area signals can be modestly more effective under controlled visual conditions. However, performance remained broadly comparable to the all-channel setup, demonstrating that restricting spatial coverage to canonical motor channels does not yield substantial gains, and extending the motor cortex \textcolor{black}{(F3, Fz, F4, P3, Pz, P4)}, including surrounding channels beyond C3, Cz, and C4, did not produce meaningful improvement. Mean accuracies in both fixation conditions ranged from 0.50 to 0.55, aligned closely with the other configurations, suggesting limited additional benefit from expanding the electrode set beyond core motor regions within this paradigm.
SVM, RF, and KNN exhibited broadly competitive and stable performance. DT and NB demonstrated slightly higher variability across subjects and conditions. No single classifier showed clear dominance, reinforcing that the bottleneck in MI performance lies in signal discriminability rather than classification strategy.
Overall, the MI paradigm achieved moderate accuracy (0.50–0.58) across all configurations and conditions, with fixation providing marginal stability improvements but no substantial accuracy gains.

 In the asynchronous hybrid MI,  all-channel configuration produces near-perfect classification performance across all subjects and classifiers. Under the fixation-off condition, mean accuracies were (Fig. \ref{fig:barplot_mi_hybrid}, bottom) 1.000 (SVM, KNN, NB), 0.999 (RF), and 0.979 (DT), with standard deviations approaching zero (SD 0.000–0.052). Under fixation-on, results were equally strong (KNN, NB: 1.000; SVM: 0.999; RF: 0.995; DT: 0.986; SD  0.000–0.044). These indicate that the asynchronous MI introduces highly discriminative, spatially distributed features that render classification virtually error-free when the full electrode montage is used. while the strict motor cortex produced a drop in performance. Under fixation-off, mean accuracies fell to 0.687–0.723 (SD: 0.139–0.166). Under fixation-on, performance was similarly limited (0.678–0.723; SD: 0.178–0.190). While expanding the electrode set to include surrounding channels, namely \textcolor{black}{F3, Fz, F4, P3, Pz, P4}, substantially recovered performance compared to the strict motor cortex setup. Under fixation-off, mean accuracies ranged from 0.792 to 0.841 (SD: 0.160–0.192). Under fixation-on, a further improvement was observed (0.828–0.853; SD: 0.174–0.187), approaching but not matching all-channel performance. 

 The asynchronous hybrid MI paradigm shows a significant improvement in classification performance, particularly for the all-channel configuration, which achieves near-perfect accuracy across all classifiers (mean 0.99–1.00) with almost zero variability. In contrast, the strict motor cortex configuration exhibits substantially lower performance (mean 0.68–0.72) with high variability (SD 0.14–0.19), suggesting limited robustness when only central electrodes are used. The extended motor cortex configuration improves performance compared to the strict setup (mean 0.82–0.85), highlighting the importance of including surrounding regions. The effect of visual fixation is minimal on the all-channel setup but slightly enhances the extended motor cortex performance. Across classifiers, SVM, RF, and KNN consistently achieve near-perfect accuracy, while DT shows minor degradation. Overall, the results demonstrate that the hybrid paradigm strongly benefits from broader spatial coverage, leading to highly reliable and stable classification performance.

 \begin{figure}[ht]
\centering
\includegraphics[width=0.5\textwidth]{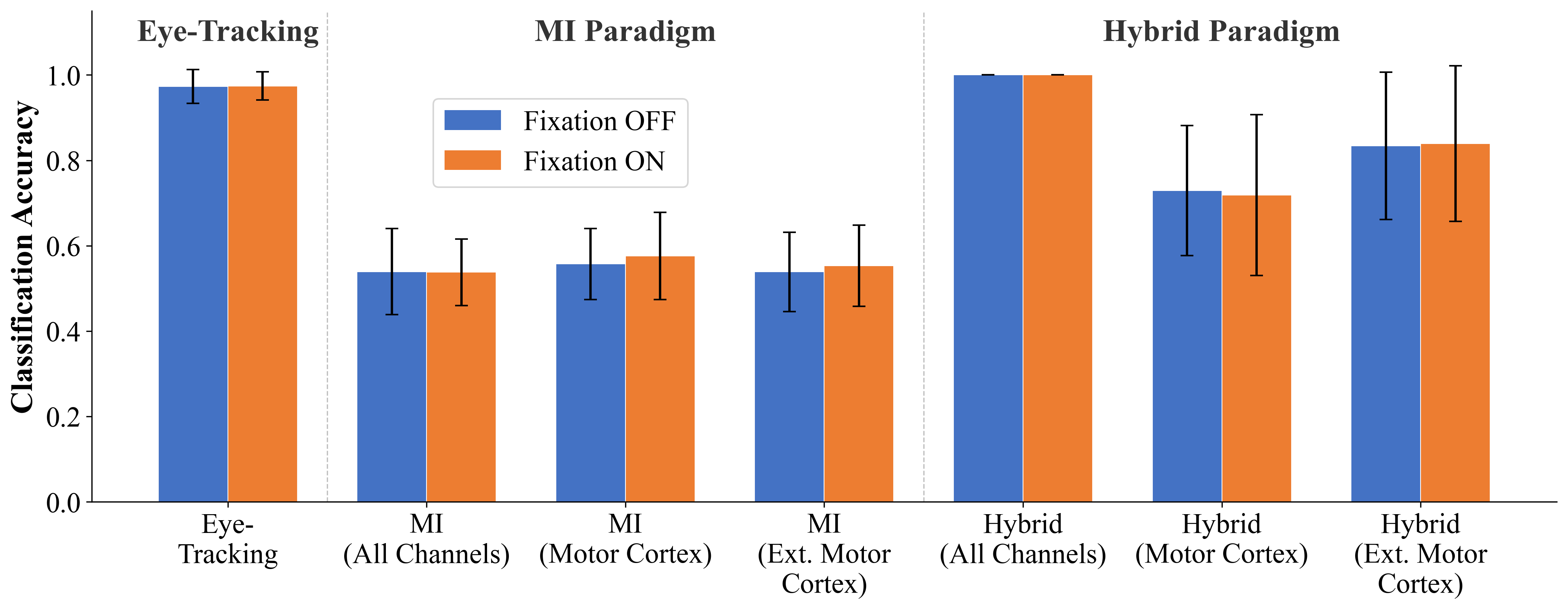}
\caption{Average accuracies (SVM) across the three paradigms: eye-tracking, MI, and hybrid MI, under fixation-off and fixation-on conditions.}
\label{fig:avg_comparison}
\end{figure}

\raggedbottom
Fig. \ref{fig:avg_comparison} summarises the average accuracy across paradigms. The 
eye-tracking component achieved approximately 75\% under both fixation conditions, while the standalone MI paradigm remained near chance level (54-58\%). In contrast, the hybrid paradigm with all channels reached near-perfect accuracy, with fixation yielding modest gains in restricted channel configurations. This hybrid approach enables single-step selection via eye tracking with asynchronous MI confirmation, eliminating repeated MI executions. A one-sided Wilcoxon signed-rank test confirmed statistically significant improvements across all classifiers ($p = 1.22 \times 10^{-4}, p < 0.001$). It should be noted that eye tracking alone serves only to identify the intended target and does not ensure reliable command execution. The proposed paradigm, therefore, incorporates MI-based confirmation, which validates the gaze-selected target before execution. The classification results reported in this section correspond to the MI confirmation stage.
\raggedbottom

\section{Discussions}
\textcolor{black}{The results highlight clear differences between the conventional MI paradigm and the asynchronous hybrid MI paradigm in terms of classification performance and robustness. The moderate performance of MI (0.50–0.58) is consistent with prior studies, where MI-based EEG classification is constrained by low signal-to-noise ratio and high inter-subject variability, leading to weak class separability \cite{an2023dual}. These limitations arise because MI-related neural patterns are distributed across the sensorimotor cortex but lack strong cross-subject consistency, reducing classifier generalisability  \cite{saha2020variability}. The marginal improvements from additional channels and visual fixation further suggest that conventional MI features, while spatially distributed, are not sufficiently discriminative on their own.} 

\textcolor{black}{In contrast, the hybrid MI paradigm shows a substantial performance improvement, achieving near-perfect accuracy (0.99–1.00) in the all-channel configuration. This aligns with previous hybrid BCI studies \cite{ o2014exploring, meena2015towards}. The drop in performance in the strict motor cortex configuration (0.68-0.72) indicates that the hybrid paradigm relies on distributed neural activity beyond the primary motor cortex, including visual and prefrontal regions recruited during asynchronous operation, consistent with observations in visual and feedback-driven BCI paradigms. 
The reduced performance with the strict motor cortex configuration suggests that the proposed paradigm engages cortical regions beyond the primary sensorimotor cortex. Unlike conventional MI, participants first process the visual target selected by gaze and then transform this information into motor imagery, recruiting additional frontal and parietal networks involved in visuospatial attention and sensorimotor integration.
The extended motor cortex configuration partially recovers performance, reinforcing the importance of broader spatial coverage in capturing the full neural signature of the hybrid task. Overall, while conventional MI remains stable but limited, the proposed hybrid paradigm significantly improves accuracy and robustness. These findings support existing literature emphasising that integrating multiple neural processes and full-channel spatial information is key to high-performance BCI systems.} Further, these findings point to broader relevance across multiple applications. For instance, gaze-based virtual keyboard interfaces for dyslexia detection \cite{meena2022detection} could adopt this hybrid framework, with MI replacing dwell time as the selection mechanism—offering a more intuitive interaction method for learning contexts where reduced complexity matters. This motivates further investigation of gaze–MI interaction for accessible learning environments.

% Additionally, when  these visual target selection done by gaze only such as virtual keyboard interfaces for dyslexia detection\cite{meena2022detection} can also extend by using this hybrid context. During the intervention phase it MI can provide the effective biomarkers. 

% \textcolor{blue}{These findings highlight the potential for developing more intuitive and accessible assistive interfaces. Such interaction may be particularly valuable in learning contexts where reduced interaction complexity is important, as demonstrated in applications supporting individuals with dyslexia \cite{meena2022detection}. This motivates further investigation of gaze–MI interaction for accessible and engaging learning environments.}

\raggedbottom
The limitations of this work include a small, demographically homogeneous sample and evaluation under controlled laboratory conditions with healthy participants, which may limit generalisability to real-world or clinical settings. Additionally, the analysis is performed offline, so real-time performance may be affected by latency, adaptation, and feedback dynamics. 
% Finally, the specific contribution of dot-motion visual feedback was not independently evaluated.
% \section{Limitation and Future Work}
% The study recruited a relatively small and demographically homogeneous sample (15 participants, predominantly aged 20–26). The near-perfect hybrid accuracies, while compelling, should be interpreted in the context of a controlled laboratory setting with healthy participants. Real-world performance with clinical populations, environmental noise, and fatigue-induced signal drift remains to be evaluated.
% Additionally, the current analysis is offline (post-hoc classification), and actual online real-time performance may differ due to processing latency, classifier adaptation requirements, and feedback dynamics. The introduction of dot-motion visual feedback as a continuous evidence accumulation mechanism is theoretically motivated, but its quantitative contribution to performance relative to the hybrid signal fusion was not independently isolated in this study.
Future work should prioritise: (1) real-time online implementation with closed-loop feedback and comparison with synchronous and asynchronous hybrid models; (2) testing on populations such as neuromuscular patients and children with dyslexia; (3) longitudinal assessment of performance stability and fatigue effects; and (4) systematic evaluation of the dot-motion feedback's contribution to decision accuracy.

\section{Conclusion}
This study investigated the impact of visual fixation and channel configuration on the performance of conventional MI and hybrid MI–eye-tracking BCIs. Results show that relying solely on motor cortex channels is insufficient, as discriminative information is distributed across broader cortical regions due to greater involvement of sensory-to-motor processing. The all-channel configuration consistently achieved superior performance, while fixation significantly improved classification accuracy and reduced inter-subject variability by stabilising attentional state during MI decoding.
The proposed asynchronous hybrid BCI paradigm substantially outperformed conventional MI, achieving near-perfect accuracy with improved cross-subject robustness. Eye-tracking integration provided complementary information, enhanced user engagement, and enabled scalable multi-command control. The hybrid approach allows direct single-step target selection via eye tracking with asynchronous MI-based confirmation, replacing the three sequential MI executions required in conventional systems. 
% . By combining
% rapid gaze-based selection with intentional MI confirmation,
% the framework reduces interaction complexity and may pro-
% vide a more accessible and cognitively efficient interface for
% dyslexic children, supporting their engagement and interac-
% tion with digital learning activities.

\textbf{Acknowledgement:}
This work was supported by grants IP/IITGN/CSE/YM/2324/05 \& ANRF/ECRG/2024/002814\\
/ENS. We thank Ashmit Chhoker for developing the UI for the initial experimental paradigm.

\bibliographystyle{IEEEtran}
\bibliography{stats}

\end{document}